\documentclass[12pt]{article}
\usepackage{amssymb}
\usepackage{graphicx}
\usepackage{subfigure}
\usepackage{indentfirst}
\usepackage[square, super, comma, sort&compress]{natbib} %style of ref
\usepackage[a4paper,left=2.5cm, right=2.5cm, top=2cm, bottom=3cm]{geometry}
\usepackage{titlesec}
\titleformat*{\section}{\large\bfseries}

\usepackage[utf8]{inputenc}
\usepackage{url}
\usepackage{hyperref}
\usepackage{amsmath}
\usepackage{amsfonts}
\usepackage{float}
\usepackage{lipsum}
\usepackage{xcolor}
\usepackage{natbib}
\usepackage{caption}

\begin{document}

%--------------------------------------------------

\begin{center}

	{\Large \textbf{Introduction of an intriguing approach for eletric}}\\ \vspace{1em}
	{\Large \textbf{current transformer on-site examining \& repairing}}\\ \vspace{3em}

	{\large Yuxuan Chen, Jing Sun, Boqi Meng*}\\ \vspace{3em}

	{Southwest Janlung Electric Technology Co., Ltd, Chongqing 400035, China}\\ \vspace{3em}

	{*Corresponding Author}\\ \vspace{3em}

\end{center}

%--------------------------------------------------

\begin{center}
    \centering
	\rule{150mm}{0.5mm}
\end{center}

\begin{abstract}

The working principle of a electric current transformer is based on electromagnetic induction, mainly composed of a closed iron core and windings. Its primary winding has relatively few turns and is connected in series with the current circuit to be measured. However, due to the frequent occurrence of full current passing through the current transformer during use, its secondary winding has relatively more turns. During use, errors may occur in the current transformer. Therefore, technical personnel can adopt the heterodyne measurement method to shield interference signals and ensure the accuracy of measurements during on-site examining and repairing of current transformers. This article mainly introduces the characteristics of transformers and their on-site examining and repairing process and errors, and describes the specific application of heterodyne measurement in on-site examining and repairing of current transformers.

\end{abstract}

\begin{center}
    \centering
	\rule{150mm}{0.5mm}
\end{center}

\vspace{5mm}

%--------------------------------------------------

\section{Characteristics of Current Transformers}
Electric current transformers are a type of power conversion device that can be classified into upright and inverted types based on their structural characteristics\cite{chen2010new,xiao2003overview}. The most commonly used in power systems is the current transformer. During its application, the secondary circuit of the current transformer is always in a closed state, and the impedance of the series coil of the measuring instrument and the protection circuit is relatively small, basically in a short-circuit working state throughout.

The role of current transformers is to convert large currents to ensure that they meet user needs and improve the safety of equipment and personnel. In recent years, a plethora of novel current transformers and associated technologies have emerged, such as the utilization of gratings to measure large currents\cite{wang2019sensitivity, zhou2021simultaneous}, optical self-compensating transformers\cite{huang2021self}, interferometric and polarimetric fiber optic current transformers\cite{muller2019temperature, bohnert2019polarimetric}, the employment of sine modulation to eliminate errors\cite{wang2021error}, and the correction of transformer errors through neural networks\cite{dybkowski2019artificial, zhao2021cnn, afrasiabi2019integration, yu2001correction, huang2022noise}, among others. Nevertheless, these optical-based transformer solutions still possess numerous vulnerabilities in practical application scenarios, such as interference from temperature\cite{lenner2019long, han2017temperature}, birefringence\cite{hu2020modeling, xu2017accurate, peng2013fiber, huang2021improving}, and vibration\cite{yu2023simultaneous, yu2022polarimetric, bohnert2002temperature}. Consequently, the industry continues to maintain that electromagnetic transformers are more suitable for power grid applications necessitating ultra-high safety factors. Of course, the author also advocates for the incubation and breakthrough of new technologies, as the ultimate objective of divergent technical routes is to benefit humanity. In order to ensure the stability of equipment operation, technical personnel must regularly check the current transformer to better grasp its working status. Once any problems are found, they must be immediately repaired or replaced. It can be seen that the role of current transformers in power systems is still very large. It can convert large currents in primary circuits into small currents in secondary circuits, making measuring instruments and protective devices smaller in size and more reasonably priced\cite{sevgili2021current,stanbury2014impact}. It is also more convenient to install in cabinets. Not only that, current transformers can also play a role in isolating high-voltage circuits. When performing primary and secondary measurements with current transformers, there is no electrical relationship between them, only magnetic contact exists. Current transformers can isolate secondary equipment from high-voltage parts, and secondary measurements are all grounded, improving the safety of equipment and personnel.

\section{Methods and precautions for on-site examining and repairing of current transformers}
\subsection{On-site examining and repairing process for current transformers}
\subsubsection{Preparation for examining and repairing}
Before starting the test, technicians should determine the equipment needed based on the actual situation on site\cite{takahashi2010field,cataliotti2009novel}. Electromagnetic transformers can expose several following problems in ultra-high voltage transmission lines: (1) Safety issues. Ultra-high voltage may cause insulation breakdown and ground short circuit. Moreover, if the secondary side is open, it will cause great harm to equipment and personnel; (2) Poor response. The iron core structure used by electromagnetic transformers limits its high-frequency response; (3) Magnetic saturation problem. The high current level of ultra-high voltage transmission lines can cause magnetic saturation of the transformer core, which seriously affects measurement accuracy; (4) Electromagnetic coils are bulky, heavy, and expensive, which is also one of the selling points claimed by electric current transformers based on optical theory to be superior to traditional transformers. At the same time, check the surrounding environment. The relative humidity should not exceed 95\%, and the error caused by electromagnetic field interference in the standard instrument should not exceed 1.2 times the error value of the current transformer being tested. The standard current transformer used for examining and repairing should have the same rated transformation ratio as the transformer to be tested, and its accuracy should be higher than that of the transformer being tested, at least two levels higher. The actual secondary load of the standard instrument should not exceed the specified upper and lower limits. If a standard instrument is used for error calibration, it is necessary to ensure that the deviation between its secondary load and the calibration certificate load does not exceed 10\%. Error measurement is also one of the more important contents in current transformer examining and repairing. Therefore, in the examining and repairing environment, the error caused by the error measurement device should not exceed 10\% of the error value of the transformer being tested.

\subsubsection{Test items}
(1) Appearance inspection. Technicians carefully check whether the appearance of the current transformer being tested is intact, whether the nameplate has recorded product numbers, date of manufacture, and rated current ratio and other signs. The port positions of the primary and secondary wiring should have symbols and signs for current wiring. The position of the wiring terminal should have a grounding sign. (2) Insulation test. By using a megohmmeter to measure the insulation resistance between each winding of the current transformer and between the winding and the ground, hand-cranked at 120 revolutions per minute, wait for the pointer to stabilize during the shaking of the megohmmeter before reading. Whether the insulation performance is qualified should be strictly followed in accordance with relevant standards and specifications \cite{annakkage2000current,mohns2017ac,saha2003some}. The primary to secondary should be greater than 1000 M$\Omega$; secondary to ground should be greater than 500 M$\Omega$; low-voltage transformers refer to test transformers greater than 5 M$\Omega$ (3) Power frequency withstand voltage test. The test voltage for primary to secondary or to ground should be operated at 85\% of the current transformer when it leaves the factory, gradually rising from a voltage close to zero, and cannot fluctuate too much. Then it drops steadily until it approaches zero voltage test. There is no odor, noise or breakdown, and maintains good insulation performance. (4) Winding polarity detection. This step requires the use of a current transformer calibrator to connect the circuit according to the comparison method. The current gradually increases until it is tested within 5\% of its rated value to determine the polarity of the transformer. (5) Error measurement. Technicians should measure at rated load, 14 load, and close to actual load respectively. The error measurement points are tested according to a fixed percentage of rated primary current, such as .1\%, 5\%, 20\%, etc., if there are special circumstances, you can continue to increase test points. The power factor of the secondary load needs to be selected according to the information marked on the transformer nameplate. During the test process, technicians should record data information well and cannot modify it at will.

\subsection{Precautions for on-site examining and repairing of current transformers When conducting on-site examining and repairing of current transformers}
First of all, technicians must ensure that one end of the secondary winding must be connected to the iron core and grounded to avoid safety accidents. Secondly, the nameplate of the transformer being tested clearly marks the level, rated power and other information. Therefore, it is necessary to ensure that the phase angle and other information will not exceed the rated range within the premise of meeting the rated power.

\section{Specific application of asynchronous frequency measurement method in on-site examining and repairing of current transformers}
\subsection{Advantages of asynchronous frequency measurement method in on-site examining and repairing of current transformers}
The main advantage of the asynchronous frequency measurement method is that it does not use 50Hz when examining and repairing the current frequency, but chooses 45Hz and 55Hz. At this time, it is easier to achieve analysis of the specified signal and minimize external interference with the current transformer. However, when using asynchronous frequency measurement method, it is necessary to have supporting software and hardware technology as support to better achieve the expected results. The output power supply of asynchronous frequency measurement method is continuously adjustable and very stable. With the support of hardware equipment, the uniform speed is faster and the accuracy is very high.

\subsection{Suppression of interference signals by asynchronous frequency measurement method}
The main purpose of applying asynchronous frequency measurement method in current transformer examining and repairing is to suppress interference signals. In traditional power systems, the application of high-speed sampling systems in power equipment is not very comprehensive. Therefore, band-pass filters are often used to suppress interference signals. 

\begin{table}[h]
\centering
\caption{The relationship between 49Hz and 51Hz measurement accuracy}
\begin{tabular}{|c|c|c|c|}
\hline
sample & 49Hz & Sample & 51Hz \\ \hline
2000   & 5.37 & 2000   & 5.39 \\ \hline
4900   & 5.01 & 5100   & 5.00 \\ \hline
9800   & 4.99 & 1020   & 4.99 \\ \hline
9800   & 4.99 & 1020   & 4.99 \\ \hline
\end{tabular}
\label{T1}
\end{table}

\begin{table}[h]
\centering
\caption{Comparison of 5 simulation results}
\begin{tabular}{|c|c|c|}
\hline
No. & 45Hz    & 55Hz    \\ \hline
1   & 5.00045 & 5.00038 \\ \hline
2   & 4.99983 & 4.99963 \\ \hline
3   & 4.99950 & 4.99982 \\ \hline
4   & 5.00082 & 4.99907 \\ \hline
5   & 4.99956 & 5.00035 \\ \hline
\end{tabular}
\label{T2}
\end{table}

If the quality of the filter cannot meet the requirements for interference suppression, the signal will become unstable. At this time, the suppression ratio of the measuring equipment can reach up to 10 times, which seriously affects the error measurement data of the current transformer. Taking 45Hz and 5Hz as examples of asynchronous frequency measurement frequencies, it is judged whether the asynchronous frequency measurement method is reasonable. The 50Hz interference separation not only has 45Hz and 55HZ. Assuming that 49Hz and 51Hz are separated from the 50Hz interference and the same asynchronous frequency measurement accuracy is achieved, then the sample size will reach as many as 9800 and 10200, which will cause two problems. The first is that too much storage resources are consumed and the requirements for the system are relatively higher. The second is that the measurement cycle is too long. Previously, a measurement took about 1 second. Now it takes about 2 seconds, which cannot meet all situational requirements. For example, some equipment related to second pulses cannot meet the measurement requirements and cannot reduce the measurement samples. Otherwise, measurement errors will occur and it will not meet actual requirements. Not only that, technicians also conducted experiments and research on the asynchronous frequency measurement accuracy of 49Hz and 51Hz. The results are shown in Table \ref{T1}. At the same time, five simulation calculations were performed on 45Hz and 55Hz. The results are shown in Table \ref{T2}. By comparison, it can be seen that using 45Hz and 55Hz as asynchronous signal measurements still meets the requirements for precision.

\subsection{Additional errors caused by asynchronous algorithms}
The calculation of additional errors through the calculation method of typical inductive sensor parameters can reduce the complexity of calculation and reduce errors. Simplification of parameters will not affect the calculation results and there is no error. This article uses the parameters of a 0.2S level current transformer as a case to test whether there is a problem with the calculation of actual power frequency errors using the heterodyne method. The rated secondary current of this current transformer is 5A, the rated capacity is 10VA, the secondary DC resistance is 0.4$\Omega$, the load is 0.4$\Omega$, and the total resistance is 0.82. When the operating current is 100\%, the excitation potential is 4V. If the current synthetic error is 0.2\%, then the secondary excitation current is 0.01A, and the excitation impedance at this time is 4002. If its in-phase component and quadrature component are the same, according to relevant formulas\cite{kumar2011time,djokic2006optically,trigo2014site}, it can be obtained that the power frequency ratio difference is -0.128\%, and the additional error of the heterodyne measurement method must be checked as 0.014\%, which is less than 1/10 of the ratio difference limit value of the 0.2S level current transformer, so it can be ignored. In addition, there are also pre-calculation situations for phase differences. Through calculation, it was found that the additional error of phase difference of heterodyne measurement method was -0.0245\%, which could also be ignored. When the operating point is 100\%, the signal-to-noise ratio of the measurement system is relatively high, and the influence of electromagnetic noise is not very large, and the error of the transformer itself is relatively small, so the benefit of heterodyne measurement method is more prominent in low-end operating currents. If the operating point is at a state of 1\% rated current, and the synthetic error is 1\%, then at this time, the secondary excitation potential is 0.04V, and the secondary excitation current is 0.005A, and its excitation impedance is 80$\Omega$. If its in-phase component and quadrature component are equal, then The error at a working point of 1\% is -0.0035\%, which can also be ignored. Through the application of heterodyne measurement method for 45Hz and 5H ratios in on-site detection of current transformers, this study investigated its suppression effect on interference abnormal signals, analyzed errors caused by this measurement method, and found that it can effectively improve performance of error measurement for current transformers, and errors produced can also be ignored.

\section{Precautions for the application of heterodyne measurement method in current transformer detection}

\subsection{Pay attention to safe operation}
When using the heterodyne measurement method to detect current transformers, attention should be paid to safe operation, strictly following operating specifications and procedures, and adopting scientific safety measures. Safe operation refers to the practice of ensuring that operations are carried out safely within a feasible range of temperature, pressure, concentration, and residence time. This can be achieved through effective design, acquisition, and installation of plant and equipment that satisfy design requirements. A licensing process and quality assurance measures must be established to ensure that these requirements are satisfied. In some industries such as machine shops, safe operation involves adhering to safety practices such as completing machine shop safety training, knowing local shop rules and emergency procedures, only operating equipment and tools that one has been specifically trained and authorized to use, following manufacturer's instruction manual or shop-developed safety procedure for each shop tool as directed by a supervisor\cite{rochlin1999safe,persson2022outside,keshishian2004duodenal}. The primary and secondary terminals of the current transformer being tested must be independent in the circuit and disconnected from other circuits, and there must be a clear disconnection point. In order to avoid the voltage in the current transformer rising to other circuits during the detection process, it is necessary to perform an electric test before starting, and then carry out the detection work after installing the grounding wire. There are several aspects to ensuring safe operation measures. First, check the isolation status of the primary wiring and operating equipment or other equipment to ensure that they are all operating independently; secondly, carefully check whether all fuses of each secondary winding are opened and whether they are in normal working condition; finally, check a half-wiring method current transformer, whether its secondary circuit is disconnected from other unrelated parallel circuits under test. Technicians should conduct tests from one side of the test circuit to avoid direct contact with the short-circuited secondary circuit, causing other equipment to malfunction. All operations in the circuit system must be carried out on the basis of ensuring safety. Therefore, managers must pay attention to on-site management to ensure the smooth progress of the entire detection work. 
\subsection{Ensure that the line status meets requirements during detection}

In current transformer detection, first of all, it is necessary to ensure that the secondary circuit of the current transformer cannot be open or short-circuited. In order to avoid such problems, technicians must first tighten the wiring of each terminal to avoid wire falling off; during examining and repairing, workers in the test area cannot move around at will, so as not to kick off the wires, which not only affects the status of the line but may also cause safety accidents; impedance generated by voltage circuits should be tested. If the test result exceeds 2M$\Omega$, it belongs to a high resistance value. Secondly, before detecting errors, ground wires must be installed properly to avoid safety accidents or equipment failures. At this time, technicians should carefully check whether the installation of ground wires is firm and reliable; ground wire installation should be within the technician’s line of sight and ensure that installation conditions meet actual requirements; if there is charged equipment around current transformers To avoid high-voltage electric shock problems, discharge first when wiring for the first time, and then connect lines; secondary windings of current transformers should use permanent grounding protection to better ensure safety. In addition, when measuring errors under secondary load of measuring transformers, since the state of secondary circuit is grounded, another end of measuring instrument cannot be grounded again. 

\section{Conclusion}
On-site detection of current transformers is a key factor in ensuring that current transformers can operate more safely and reliably during use. On-site detection can directly check performance parameters such as voltage withstand capability, insulation performance and volt-ampere characteristics of current transformers, laying a foundation for safe operation of equipment and safe and stable transmission of electric energy. In actual operation process, technicians must pay attention to safety protection work and ensure insulation performance of all equipment. Complete test tasks on premise of ensuring safety and make more contributions to development of power enterprises.

\clearpage

%--------------------------------------------------

\footnotesize

\bibliographystyle{ieeetr}
\bibliography{references.bib}

\end{document}